\documentclass[usegraphicx,usenatbib]{mn2e}
\usepackage{longtable}
\usepackage{graphicx}
\usepackage{graphics}
\usepackage{keyval}
\usepackage{trig}
\usepackage{graphicx}
\usepackage{dcolumn}
\usepackage{bm}
\usepackage{subfigure}
\def\beq{\begin{equation}}
\def\eeq{\end{equation}}
\def\bey{\begin{eqnarray}}
\def\eey{\end{eqnarray}}
\def\msun{M_\odot}
\def\lsun{L_\odot}
\def\kms{\, {\rm km \, s}^{-1} }

\def\mnras{MNRAS}
\def\apj{ApJ}

\def\apjs{ApJ}
\def\apjl{ApJ}
\def\araa{ARAA}
	
\def\aap{A \& A}
\def\aj{AJ}

\def\aap{Astron. Astrophys.}
\begin{document}

\title{Resolving the timing problem of the globular clusters orbiting the Fornax dwarf galaxy}

\author[G. W. Angus, A. Diaferio]{G. W. Angus$^{1,2}$\footnote{email:angus@ph.unito.it}, A. Diaferio$^{1,2}$ \\
$^{1}$Dipartimento di Fisica Generale ``Amedeo Avogadro", Universit\`a degli studi di Torino, Via P. Giuria 1, I-10125, Torino, Italy \\
$^{2}$Istituto Nazionale di Fisica Nucleare (INFN), Sezione di Torino, Torino, Italy\\}

\date{\today}
\maketitle
\begin{abstract}
We re-investigate the old problem of the survival of the five globular clusters orbiting the Fornax dwarf galaxy in both standard and modified Newtonian dynamics. For the first time in the history of the topic, we use accurate mass models for the Fornax dwarf, obtained through Jeans modelling of the recently published line of sight velocity dispersion data, and we are also not resigned to circular orbits for the globular clusters. Previously conceived problems stem from fixing the starting distances of the globulars to be less than half the tidal radius. We relax this constraint since there is absolutely no evidence for it and show that the dark matter paradigm, with either cusped or cored dark matter profiles, has no trouble sustaining the orbits of the two least massive globular clusters for a Hubble time almost regardless of their initial distance from Fornax. The three most massive globulars can remain in orbit as long as their starting distances are marginally outside the tidal radius. The outlook for modified Newtonian dynamics is also not nearly as bleak as previously reported. Although dynamical friction inside the tidal radius is far stronger in MOND, outside dynamical friction is negligible due to the absence of stars. This allows highly radial orbits to survive, but more importantly circular orbits at distances more than 85\% of Fornax's tidal radius to survive indefinitely. The probability of the globular clusters being on circular orbits at this distance compared with their current projected distances is discussed and shown to be plausible. Finally, if we ignore the presence of the most massive globular (giving it a large line of sight distance) we demonstrate that the remaining four globulars can survive within the tidal radius for the Hubble time with perfectly sensible orbits.
\end{abstract}

\section{Introduction}
\protect\label{sec:intr}

The Fornax dwarf galaxy is the largest of the dwarf spheroidal galaxies of the Milky Way. Its integrated stellar luminosity is $L_v=(1.58\pm0.16)\times 10^7\lsun$ (\citealt{mateo98}), but it has retained virtually no mass ($<5000\msun$) in gas (\citealt{mateo91}), which makes pinpointing its radial velocity less accurate. Regardless, the best estimates (\citealt{mateo91}) suggest a Heliocentric distance and advancing radial velocity of $138\pm8~kpc$ and $53\pm3\kms$ respectively. Fornax has a well measured surface density profile enabling the deduction of its King model parameters which define the number density of stars from the centre to the tidal radius, $R_t=2.08\pm0.18$~$kpc$. This is complemented by the concentration parameter $\log_{10}(R_t/R_c)=0.72$ (\citealt{irwinhatz}). 

Stellar population synthesis models tell us that for the ages and metallicities of stars in Fornax, the ratio of mass to luminosity should be of order unity (\citealt{kroupa01}), however, the first puzzle of Fornax, as with the other dwarfs, is that relating the random motions of the stars to their implied mass from their integrated luminosity leaves an incongruity: the gravity $GM_{\star}(r)r^{-2}$ is simply insufficient. There is an ``acceleration deficit".

Since we assume the measurements are correct, either gravity is not Newtonian, allowing the acceleration to be boosted according to some algorithm, or there is some dark matter (DM) which provides a similar service. What this means in detail is discussed in \S\ref{sec:dm} and \S\ref{sec:mond}, nevertheless, the observations of Fornax have recently become so detailed (\citealt{walker07}) that more than two thousand member stars can be sorted into different projected radii bins giving the line of sight (los) velocity dispersion (VD) as a function of projected radius. This basically fixes the allowed mass density, luminous or dark, at all radii where stars exist.

The second puzzle is far more subtle. In orbit, there are at least 5 well resolved globular clusters (GCs), with important parameters given in Table 1. The GC masses range between $0.37\times 10^5\msun$ and $3.63\times 10^5\msun$, the projected distances are from 0.24 to $1.6~kpc$ and their los relative velocities with respect to Fornax are at most $8.7\pm 3.6\kms$.

\begin{table*} 
\begin{tabular}{|c|c|c|c|c|c|}
\hline
GC& Mass $10^5\msun$ & $D_p$ [kpc] & $V_{los}$ [$\kms$]& Smallest Pericentre [kpc] & Velocity at Pericentre [$\kms$]\\
\hline
1 & 0.37 &  1.60  & ...   &...  & ...\\
2 & 1.82 &  1.05  & -1.2 $\pm$ 4.6  &1.20 & 34\\
3 & 3.63 &  0.43  & 7.1 $\pm$ 3.9 &1.45 & 31\\
4 & 1.32 &  0.24  & 5.9 $\pm$ 3.4   &1.05 & 36\\
5 & 1.78 &  1.43  & 8.7 $\pm$ 3.6   &1.20 & 34\\
\hline
\end{tabular} 
\medskip 
\caption{Parameters for the 5 GCs. Masses and projected distance ($D_p$) come from Mackey \& Gilmore (2003) and los relative velocities ($V_{los}$) come from Mateo et al. (1991). The pericentres and velocities at pericentre correspond to Fig~\ref{fig:gcs} and are the smallest pericentres that can survive a Hubble time in the MOND model.}
\end{table*} 
As observers, we have no knowledge of the third spatial dimension (depth), nor the motion in the plane of the sky. What we do know, is that if we assume these GCs are a representative sample, then there is very little chance that all 5 have los distances from Fornax that are considerably greater than the maximum projected distance, although one might. Therefore, it is a safe bet that all 5 currently spend the majority of their orbits at projected distances less than $1.6~kpc$ and their orbits must reflect this.

There is more freedom in the orbital velocity, since only one of the three coordinates are measured (los, not the two dimensions in the plane of the sky) and with far poorer precision than the projected radius. Three out of four of the GCs have los velocities of between 7 and $9~\kms$ w.r.t Fornax, but there is no real statistical significance attached to this.

Several studies (\citealt{read06,goerdt06,sanchez06,strigari06,inoue09}) have suggested that the mere presence of GCs near Fornax represents a fundamental problem for concordance cosmology (\citealt{spergel07}) or the modified Newtonian dynamics of \cite{milgrom83a}. They claim that the timescale over which the orbital angular momentum of these GCs is drained by the background of stars and DM, through which they orbit, is a fraction of the age of the Universe. Despite this, there is no bright nucleus of stars at the centre as there would be if other GCs had decayed (\citealt{tremaine76}). This process which bleeds the orbital angular momentum is called dynamical friction (DF) and, in the classical sense, is given by (\citealt{bt08})

\bey
\protect\label{eqn:df}
\nonumber a_{df}(r)={4\pi \ln \Lambda G^2 \rho(r) M_{GC} \over V_c(r)^2}\times\\
\left[erf \left({V_c(r) \over \sqrt{2} \sigma } \right) - \sqrt{2 \over \pi} {V_c(r) \over \sigma} \exp\left(-{V_c(r)^2 \over 2 \sigma^2} \right) \right]
\eey
where $\ln \Lambda$ is the Coulomb logarithm and is taken to be approximately 3. $V_c(r)$ is the circular velocity of Fornax, $G=4.4\times10^{-6}\kms kpc^{-1}\msun^{-1}$ is Newton's constant, $M_{GC}$ is the mass of the GC and $\sigma$ is the measured VD of Fornax which is roughly $11~\kms$. We have knowledge of all relevant parameters in the Newtonian case once a density profile, $\rho(r)$, of luminous plus DM has been fitted to the losVD.

\section{dark matter models}
\begin{figure}
\includegraphics[angle=0,width=7cm]{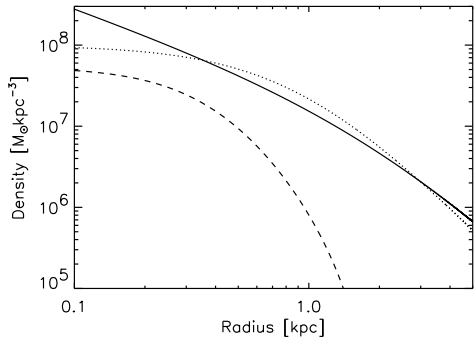}
\caption{Mass density profiles for cusped DM (solid), cored DM (dotted) and MOND (dashed).}\label{fig:rho}
\end{figure}

\protect\label{sec:dm}
We use the procedure of \cite{aftcz} and \cite{angus08} to match the losVD profile of Fornax by solving the Jeans equation assuming a cusped NFW profile (\citealt{nfw97}) and the same cored DM profile employed by (\citealt{read06,goerdt06,sanchez06,inoue09}; hereafter R06, G06, S06 and I09 respectively). We neglect the density corresponding to the stars, since it has no direct bearing on the dynamics and can be simply subtracted if one wants to infer the exact density of DM, which is not our goal here. The fits to the losVD are shown in Fig~\ref{fig:losvd} (solid line) and the NFW density profile $\left(\rho(r)={\rho_o r_s^3 \over r(r+r_s)^2}\right)$ is plotted in Fig~\ref{fig:rho} with parameters $r_s=2.5~kpc$, $\rho_o=1.2\times10^7\msun kpc^{-3}$, a concentration of $c={r_{200} \over r_s}=7.0$ and the velocity distribution of stars is isotropic ($\beta(r) \equiv 1-{\sigma^2_{\theta}\over \sigma^2_r}=0.0$). The enclosed mass of stars and DM is given in Fig~\ref{fig:mass}.

It is this constraint, and not the mere existence of GCs in orbit that must fix the DM halo since the NFW profile works perfectly well in this case and it is not possible to have a better match to the data, regardless of whether $7.0$ is a sensible concentration parameter for this size of halo, nor if the DM density is suspiciously low for a cold DM particle. Nevertheless, to compare with the work of R06, G06 and I09 we take the cored density profile they use ($\rho(r)=\rho_o \left[1+({r \over r_s})^2\right]^{-1.5}$; where $\rho_o=10^8\msun kpc^{-3}$ and $r_s=0.91~kpc$) and attempt to match the losVD. It is not possible to obtain a good match even with a variable anisotropy and to have a satisfactory fit one needs highly radial orbits, $\beta(r)=0.46$. A more aesthetic cored profile has parameters $\rho_o=2.2\times10^8\msun kpc^{-3}$, $r_s=0.5~kpc$ and isotropic velocity anisotropy ($\beta(r)=0$), but we only mention it here. In any case, as one sees from Fig~\ref{fig:df}, which plots the deceleration from dynamical friction given by Eq~\ref{eqn:df}, the two DM profiles are indistinguishable beyond $0.3~kpc$ and so we discuss them jointly hereafter.

Knowledge of the mass profile also tells us the circular velocity profile ($V_c(r)^2={G \over r} \int_0^r \rho(\tilde{r}).4\pi \tilde{r}^2 d\tilde{r}$), thus, we have all the pertinent information that allows us to follow the orbits of the GCs.

This case where the Fornax dwarf is dominated by the existence of DM particles, as opposed to modified gravity, requires very little discussion because it has absolutely no problem with sustaining the orbits of GCs for a Hubble time. 

\begin{figure}
\includegraphics[angle=0,width=7cm]{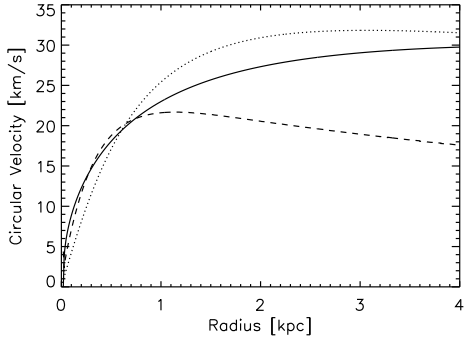}
\caption{Circular velocity profiles for cusped DM (solid), cored DM (dotted) and MOND (dashed).}\label{fig:vc}
\end{figure}

Our procedure is to numerically solve the equations of motion for a GC with a starting $x$ and $y$ position and velocity w.r.t Fornax (we use zero $z$ distance - line of sight - and velocity for ease). The acceleration due to gravity (Fig~\ref{fig:grav}) is computed from the circular speed (Fig~\ref{fig:vc}), $V_c(r)^2r^{-1}$, and the deceleration from dynamical friction is given by Eq~\ref{eqn:df} and shown in Fig~\ref{fig:df}. Combining all this allows us to follow the orbit of the GC using simple Eulerian time-stepping such that for the $x$ coordinate

\bey
V_{x_{i+1}}&=&V_{x_{i}}-dt \left[{V_c^2(r_i) \over r_i}{x_i \over r_i}+a_{df}(r_i){V_{x_i} \over |V|} \right], \nonumber \\
x_{i+1}&=&x_{i}+dt V_{x_{i+1}}
\eey
where the time step $dt=0.01~Myr$, is less than 1 per cent of the dynamical time at $1~kpc$. To update the $y$ coordinate we simply swap the $x$ subscripts for $y$.

Although in the DM scenario we use only circular orbits, the use of both $x$ and $y$ coordinates (and not merely $r$) allows us to use more exotic orbits like highly elliptical ones.

\begin{figure}
\includegraphics[angle=0,width=7cm]{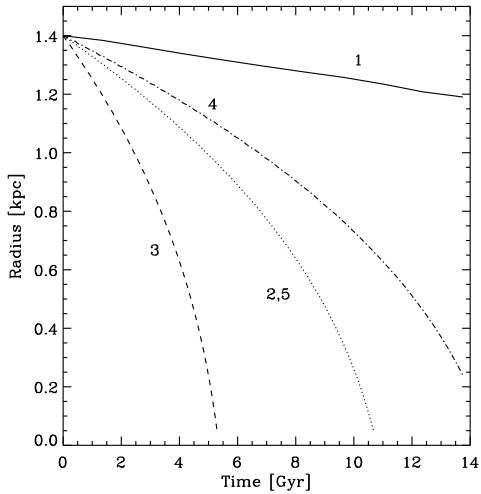}
\caption{The orbital decay of the 5 GCs for the DM profile beginning with circular orbits at $1.4~kpc$.}\label{fig:idf}
\end{figure}

In Fig~\ref{fig:idf} we show the decay of the GCs from a distance of $1.4~kpc$ from Fornax which is close to the maximum observed projected distance. This starting distance is nothing more than a sensible, assumed current 3-d distance for the 5 GCs to illustrate a point. Clearly, GC 3 decays to the centre in less than $5~Gyrs$, but it takes about $10~Gyrs$ for GCs 2 and 5, which is a large fraction of the Hubble time. GCs 1 and 4 can survive in orbit for a Hubble time if they begin at $1.4~kpc$. So really, if the GCs began their orbits at $1.4~kpc$ then only one of the 5 GCs has a problem surviving, although actually they need to do slightly more than survive because the current projected distances are around $1~kpc$ from Fornax.

Fig~\ref{fig:dfn} shows orbits that start at projected distances that decay towards 1kpc from Fornax (the average current projected distance). GC 3, the most massive one, must originally fall from outside Fornax's tidal radius to currently be near $1~kpc$, which is absolutely no problem. In any case, as we shall see in the MOND section, non-circular motions can slow the infall rate by allowing the GC to spend significant portions of the orbit in weak DF regions. Furthermore, since the other 4 GCs are not at all vulnerable to DF, the GC that {\it is} becomes a statistic of one and very little significance can be attached to it. For instance, this single GC could easily have a los distance of several $kpc$ and may not even be bound to Fornax. Therefore, the existence of these GCs cannot be used to gainsay the type of DM halo surrounding Fornax: whether it be cored or cusped, the orbits are perfectly stable to DF.

The reason it has been mistaken as a problem for the concordance cosmology is due to the initial conditions chosen by the three main previous studies (R06, G06, S06) and more recently by I09. Without justification, \cite{read06} begin all simulations from a starting distance of $1~kpc$, \cite{goerdt06} increase this slightly to $1.1~kpc$ for their maximum starting distance (but often much lower) and \cite{sanchez06} decrease it again to $0.9~kpc$, but often use $0.6~kpc$ to emphasise their point. I09 begin most simulations at $0.6~kpc$, and some at $1~kpc$.

Sticking to the facts, there is no evidence that the GCs were at a distance of 1kpc a Hubble time ago. The only clues are that there are now 5 GCs with a maximum projected distance of $1.6~kpc$. It is only the statistical significance of observing 5 within a projected distance of $1.6~kpc$ that makes us surmise that the los distances are not significantly larger than the projected distances. Nevertheless, there is nothing to say they could not have decayed from a larger projected distance to their current one. What these authors have shown conclusively is that starting with circular orbits from less than $1~kpc$ is not a favourable method to create the current positions.

\section{Dynamical Friction in MOND}
\protect\label{sec:mond}

When \cite{milgrom95} first studied the dynamics of the dwarf galaxies in MOND there were only single value (central) losVDs and therefore only a rough guide to the internal acceleration of each dwarf was required to predict the M/L because other unknowns such as the velocity anisotropy, luminosity, distance amongst the still unknown impact of interloper stars made it a rather inaccurate business. In that case, the dwarf was deemed to be either isolated, wherein its internal gravity was stronger than the external field of the Milky Way, or dominated by the external field of the Milky Way. Now that the velocity dispersion of Fornax along the los is measured fairly accurately as a function of radius, it is important to include the MW in the MOND equation since it is a non-linear gravity theory.

Therefore, we must numerically solve for the gravity at all radii from Fornax using the equation

\beq
\protect\label{eqn:mond}
g(r)\mu(x)=GM_{\star}(r)r^{-2},
\eeq
where $\mu(x)={x \over \sqrt{1+x^2}}$, $x={g(r)+g_{ex} \over a_o}$ and $g_{ex}={V_{c,MW}^2 \over R_{MW}}={(170\kms)^2 \over 138kpc}=210(\kms)^2 kpc^{-1}$ (see \citealt{mcgaugh08} for a detailed, MOND-inspired Milky Way model and \citealt{xue08} for the evidence suggesting the flat rotation speed of the Milky Way is $170~\kms$). Notice that in the absence of the external field, $g_{ex}$ from the Milky Way, we would regain the standard MOND relation, $g= \sqrt{g_na_o}$, for $g << a_o$.

\begin{figure}
\includegraphics[angle=0,width=7cm]{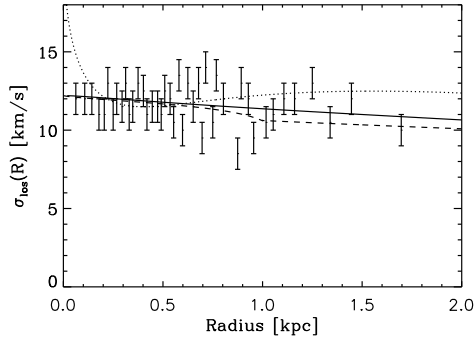}
\caption{The data points are the losVD as measured by Walker et al. (2007). The solid line is the Newtonian fit for a cusped NFW profile with $r_s=2.5~kpc$ and $\rho_o=1.2\times 10^7\msun kpc^{-3}$ and isotropic orbits $\beta \equiv 1-{\sigma^2_{\theta}\over \sigma^2_r}=0.0$. The dotted line takes the parameters used in the analyses by Read et al. (2006), Goerdt et al. (2006) and Inoue (2009) with $\beta(r)=0.46$ to facilitate better agreement between the observations. The dashed line is the MOND fit using only the luminous matter with M/L=1.4, a velocity anisotropy that varies as $\beta=-0.36\left({r^2 \over r^2+(0.3kpc)^2}\right)$ and external field strength $g_{ex,MW}={(170\kms)^2 \over 138kpc}$.}
\label{fig:losvd}
\end{figure}

\begin{figure}
\includegraphics[angle=0,width=7cm]{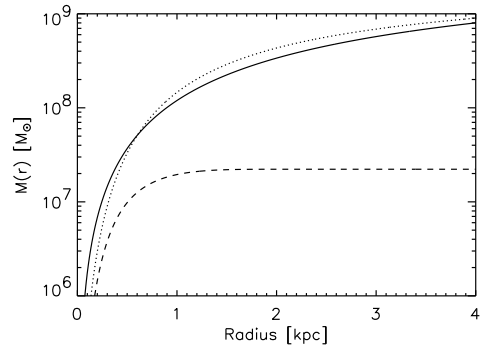}
\caption{Enclosed mass profiles for cusped DM (solid), cored DM (dotted) and MOND (dashed).}\label{fig:mass}
\end{figure}

$M_{\star}(r)$ (plotted in Fig~\ref{fig:mass}) is the enclosed mass of Fornax defined by the King luminosity profile with parameters mentioned in the introduction using a mass to light ratio (M/L) of 1.4 which is used to match the losVD profile in MOND (Fig~\ref{fig:losvd} and \citealt{angus08}).

\begin{figure}
\includegraphics[angle=0,width=7cm]{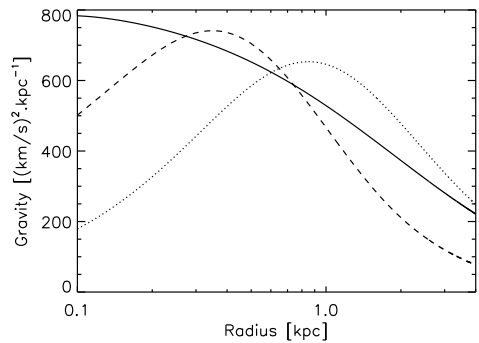}
\caption{Gravity as a function of distance for the cusped DM profile (solid), cored DM (dotted) and MOND (dashed). Recall that $a_o$ in these units is 3600.}\label{fig:grav}
\end{figure}

The gravity $g(r)$ is plotted in Fig~\ref{fig:grav} (dashed line) and is always considerably smaller than $a_o=3600(\kms)^2 kpc^{-1}=1.2\times10^{-10}ms^{-2}$. Taking this one more step, at all points in Fornax, the gravity of the GC is increased by the factor ${a_o \over g(r)}$. Notice the distinction here between Fornax and the GC's gravity: Fornax is only mildly affected by an external field (that of the MW) and so we use Eq~\ref{eqn:mond} to infer Fornax's gravity as a function of radius. The gravity of the GC is dominated by the external field of Fornax, $g(r)$, and so we can make the approximation that the gravity of the GC is boosted (from its Newtonian value) by the factor ${a_o \over g(r)}$.

With this in mind, recall that DF in the Newtonian sense is given by Eq~\ref{eqn:df}. Analytically, this equation has previously been altered for MOND (\citealt{ciotti04}) by noticing the $G^2$ term must be corrected to account for the MOND effect at the GC's position i.e. in Eq~\ref{eqn:df} we must replace $G^2 \rightarrow G^2({a_o \over g(r)})^2$ (see \citealt{ciotti04} and also \citealt{sanchez06}).

It should be recognised that this analytical description of dynamical friction in MOND is quite poorly motivated since this gravitational boost factor $({a_o \over g(r)})^2$ varies with position inside Fornax, since $g(r)$ varies with radius; actually by a factor of four from its maximum to the tidal radius. Furthermore, out to a distance $\sqrt{{GM_{GC} \over a_o}}$ from the GC itself, the gravity is fully Newtonian: this is around $20~pc$ for GC 3. The DF formula used does not explicitly take these factors into account, calculating the boost factor at the GC's position, instead of accounting for the boost factor along the orbit of the star that is scattered by the GC. Therefore, it is a slightly uncertain upper limit and ideally we would prefer N-body simulations of GC orbits in Fornax.

\cite{nipoti08} studied DF of bars in galaxies with high resolution N-body simulations in MOND (see also \citealt{tc08}), and found DF is significantly stronger in that special case. \cite{sanchez06} investigated this exact issue we study and showed that circular orbits beginning at $0.9~kpc$ (less than half the tidal radius of Fornax and significantly less than 3 of the projected distances of the GCs) are rather short lived and claimed this is a critical issue for MOND. As discussed above, this is not a rigorous enough treatment of the problem to label this as a problem for the MOND paradigm. We make a more definitive statement in Fig~\ref{fig:dfm} for circular orbits beginning slightly within the tidal radius of Fornax (at $1.8~kpc$). In particular, notice that the survival time for 4 GCs from this distance is greater than $8~Gyrs$ and from $1.9~kpc$ the decay time is infinite (discussed further in \S\ref{sec:circrt}) because the density of stars approaches zero. Nevertheless, circular orbits beginning at $\sim 1.6~kpc$ a Hubble time ago are most definitely problematic.

\subsection{Radial Orbits}
\protect\label{sec:radial}
We have established that circular orbits within $1.6~kpc$ for the three more massive GCs have short survival time w.r.t the Hubble time, but are there other orbits that enter the tidal radius that can survive for significantly longer?

Assuming the GCs do indeed enter within the tidal radius of Fornax, the only scenario that can prevent them decaying in a period much shorter than the Hubble time is to have highly radial orbits that pericentre well outside the core radius, $R_c=0.4~kpc$, but inside the tidal radius, $R_t=2.08kpc$. These orbits represent the most penetrating orbits that can survive a Hubble time. 

Given that the orbits extend out to several $kpc$, it is important to show they remain within the inner Lagrange point in the GC, MW, Fornax system. A comprehensive analysis of Roche lobes was developed for MOND in \cite{zhaot06}, where they show a GC would have to be at a distance $D_L \approx R_{MW}\left(\xi {M_* \over M_{MW}} \right)^{1/3}$. Using a Milky Way distance, $R_{MW}$, mass $M_{MW}=5\times10^{10}\msun$ like before (\citealt{mcgaugh08} and references therein) and $\xi=1$ recommended by \cite{zhaot06} for the deep MOND regime, this gives $D_L \approx 12~kpc$ which is significantly greater than the largest apocentre reached.

The orbital parameters corresponding to these orbits are given in Table 1, including the pericentric distance and tangential velocity (angular momentum) at pericentre, but only for the four most massive GCs, since the lightest one is not problematic. Fig~\ref{fig:gcs} shows the orbital decay of the GCs for the orbits with the lowest possible pericentre that can survive a Hubble time. Since we have only three out of six of the phase space coordinates we cannot with absolute certainty rule out the radial orbits presented in Fig~\ref{fig:gcs}. In spite of this, the fact remains that the 5 GCs currently have projected distances less than $1.6~kpc$ (and the 4 most massive and significant GCs are within $1.4~kpc$). Therefore, the orbits of the GCs must not only survive a Hubble time, but also reflect the current positions with reasonable probability.

To help attach some statistical significance to this problem, we force the orbits of the GCs to be inclined so that the angular momentum is entirely in the plane of the sky. This is not particularly convoluted because if we take the eight or more dwarf satellites of the Milky Way as examples (see \citealt{lb83,dong08}), their orbits are not random, but follow two great circles around the Milky Way presumably because they were part of two smaller groups of galaxies that entered the Milky Way at late times.

Therefore, we are looking to find orbits that have a high probability of being observed within $1.4~kpc$ of Fornax andwe calculate this probability simply by counting how many time steps in the orbit integration were spent within a projected radius less than $1.4~kpc$, we found that the 4 largest GCs spend $\approx$45\% of each orbit within $1.4~kpc$ of the centre of Fornax. This corresponds to a probability of at most $0.45^4 \approx 0.04$ for these orbits averaged over the Hubble time, which seems unfeasible.

However, the special thing about these orbits is that within the last two or three Gyrs the GCs have spent all of their orbit inside $1.4~kpc$. As can be seen in Fig \ref{fig:gcs} the orbits are increasingly circularised with time and this amplifies the probability of being observed within $1.4~kpc$. This time period can be further increased if the GCs were captured at some fraction of the Hubble time, since the orbits need not be as extreme. It must be stated that these orbital parameters are fine-tuned, but it is still important to be aware such orbits exist, since sadly we have no facts about their history.

\subsection{Circular Orbits at the Tidal Radius}
\protect\label{sec:circrt}

\begin{figure}
\includegraphics[angle=0,width=7cm]{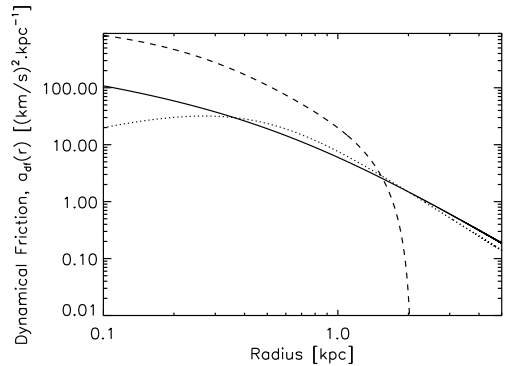}
\caption{Dynamical Friction as a function of distance from Fornax and GC-3 for the cusped DM profile (solid), cored DM (dotted) and MOND (dashed). Recall that $a_o$ in theses units is 3600.}\label{fig:df}
\end{figure}

A crucial point is that if the GCs have circular orbits beyond the tidal radius (in fact 85\% of $R_t \rightarrow$ $1.9~kpc$) the orbits are permanently stable. As one can see from Fig~\ref{fig:df} or Eq~\ref{eqn:df}, the DF deceleration tends to zero around $2~kpc$ as the density of stars from the King model tends to zero at the tidal radius. Here, there are no stars to scatter, whereas in the Newtonian framework the omnipresent DM particles are still abundant. In actual fact, we have no dynamical tracers beyond the stellar tidal radius, so the DM might also be truncated.

One way to estimate the probability of such a scenario is to think of the orbital time spent outside the maximum projected distance for a circular orbit at $D_{p,max}=2~kpc$, where the circular velocity is $20~\kms$. If we again accept the orbit is inclined so that all angular momentum is in the plane of the sky, the relationship for the projected position of such an orbit is simply

\beq
\protect\label{eqn:circ}
D_p=2\sin\left({t \over 100}\right) [kpc]
\eeq
where the angular frequency ${1 \over 100Myr}$ is $V_c D_{p,max}^{-1}$. It can easily be seen that for $t>77.5~Myr$, $D_p>1.4~kpc$ which is the maximum projected distance of the 4 largest GCs. What we want to know is whether observing the 4 GCs within $1.4~kpc$ is a contrivance if they have a circular orbit beyond $2~kpc$. Since 77.5\% of the time a GC on a circular orbit at $2~kpc$ is within $1.4~kpc$, the probability of observing it within $1.4~kpc$ is $\left({77.5 \over 100}\right)$. Therefore, the probability of observing all four within $1.4~kpc$ is $\left({77.5 \over 100}\right)^4=0.36$, which one could argue makes this a perfectly plausible scenario.

\subsection{GC 3}

\begin{figure}
\includegraphics[angle=0,width=7cm]{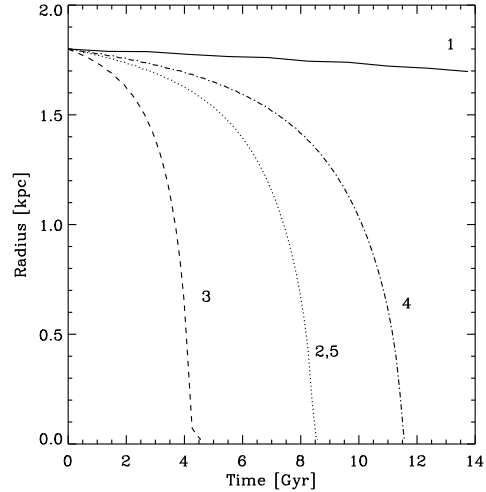}
\caption{The orbital decay of the 5 GCs in MOND beginning with circular orbits just inside the tidal radius ($1.8~kpc$).}\label{fig:dfm}
\end{figure}

\begin{figure*}
\def\subfigtopskip{0pt} 
\def\subfigbottomskip{4pt}
\def\subfigcapskip{1pt}
\centering

\begin{tabular}{cc}
\subfigure{\label{fig:rad25}
\includegraphics[angle=0,width=7.0cm]{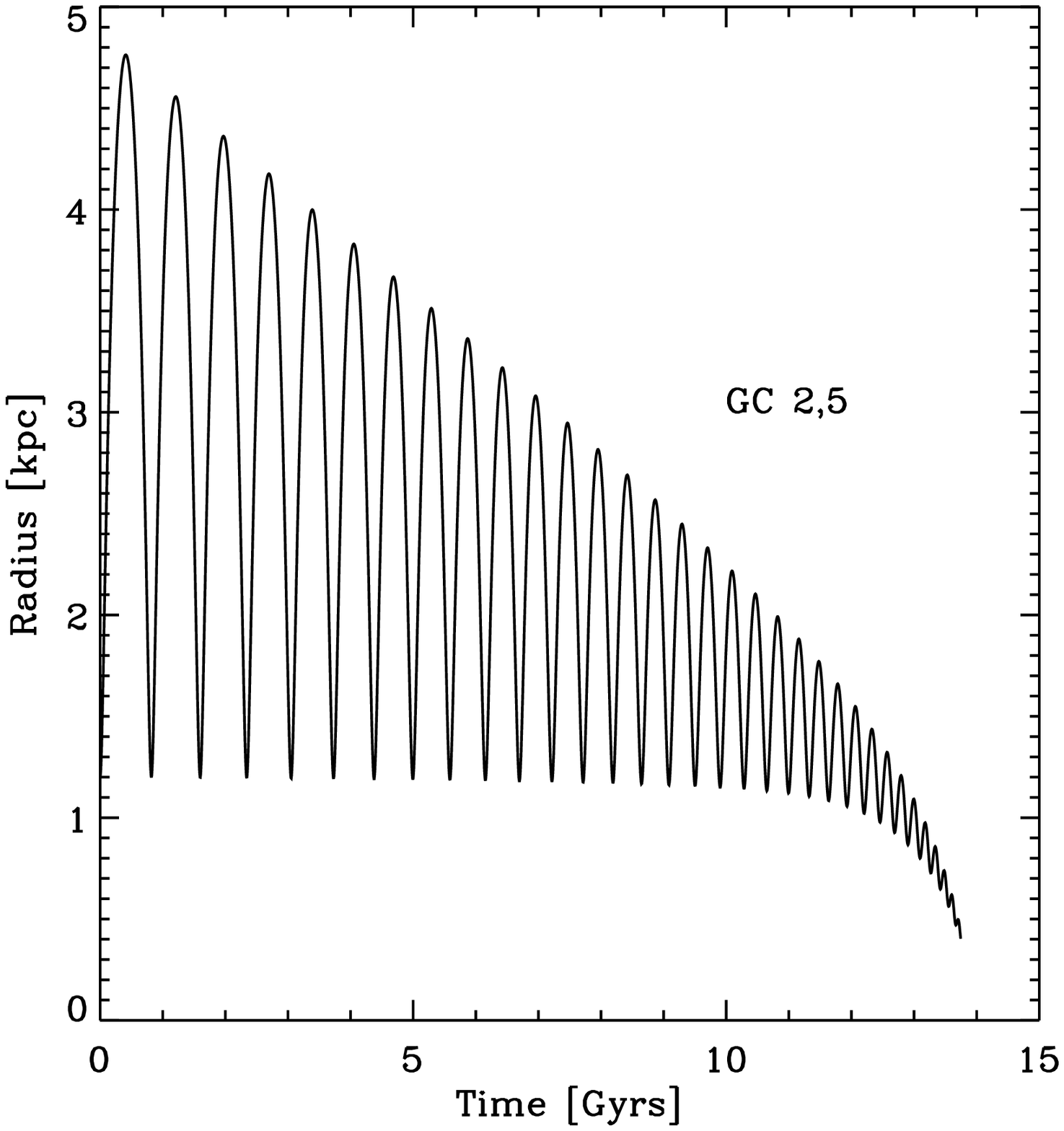}
}
&
\subfigure{\label{fig:spi25}
\includegraphics[angle=0,width=7.0cm]{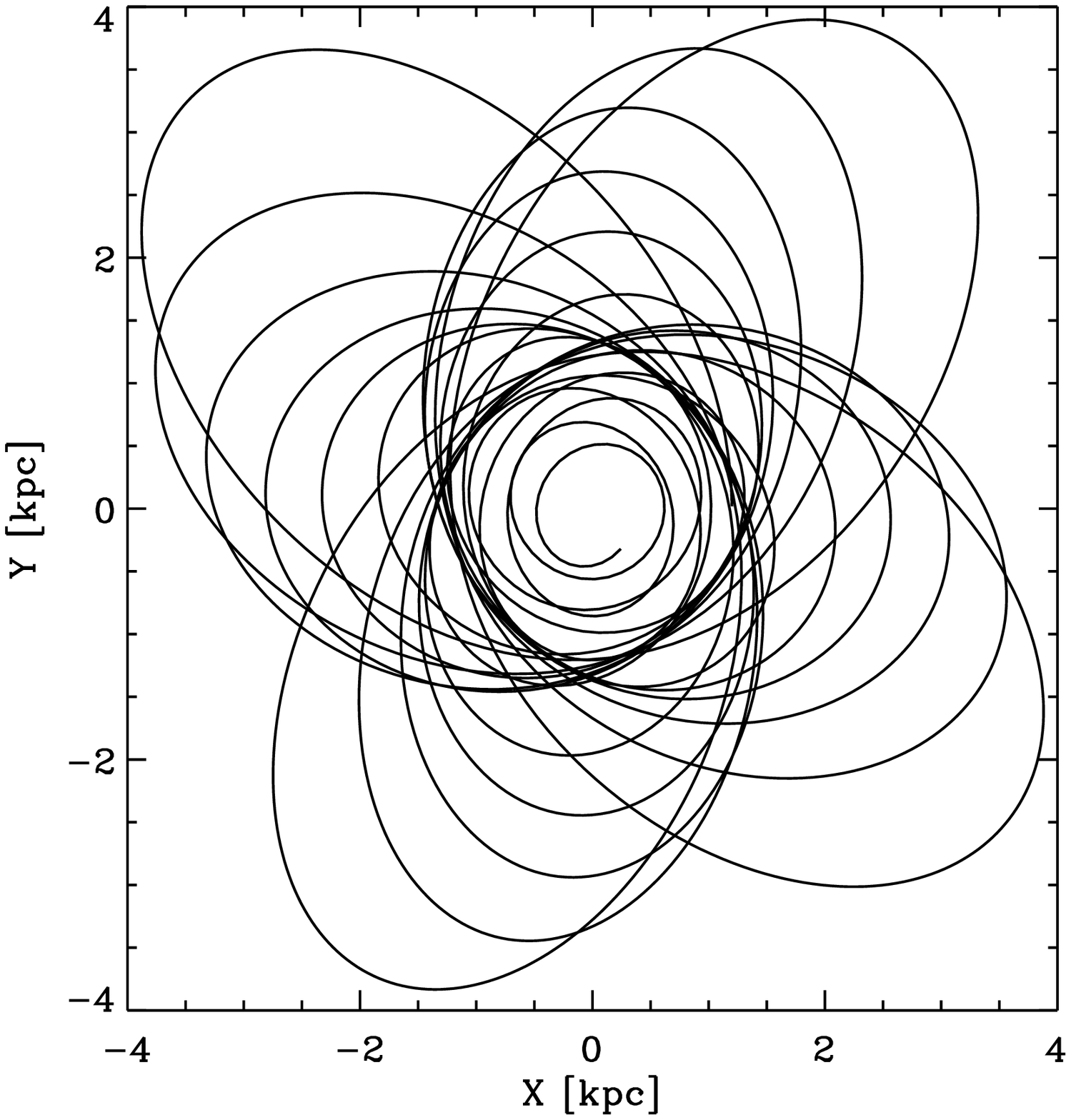}
}
\\
\subfigure{\label{fig:rad3}
\includegraphics[angle=0,width=7.0cm]{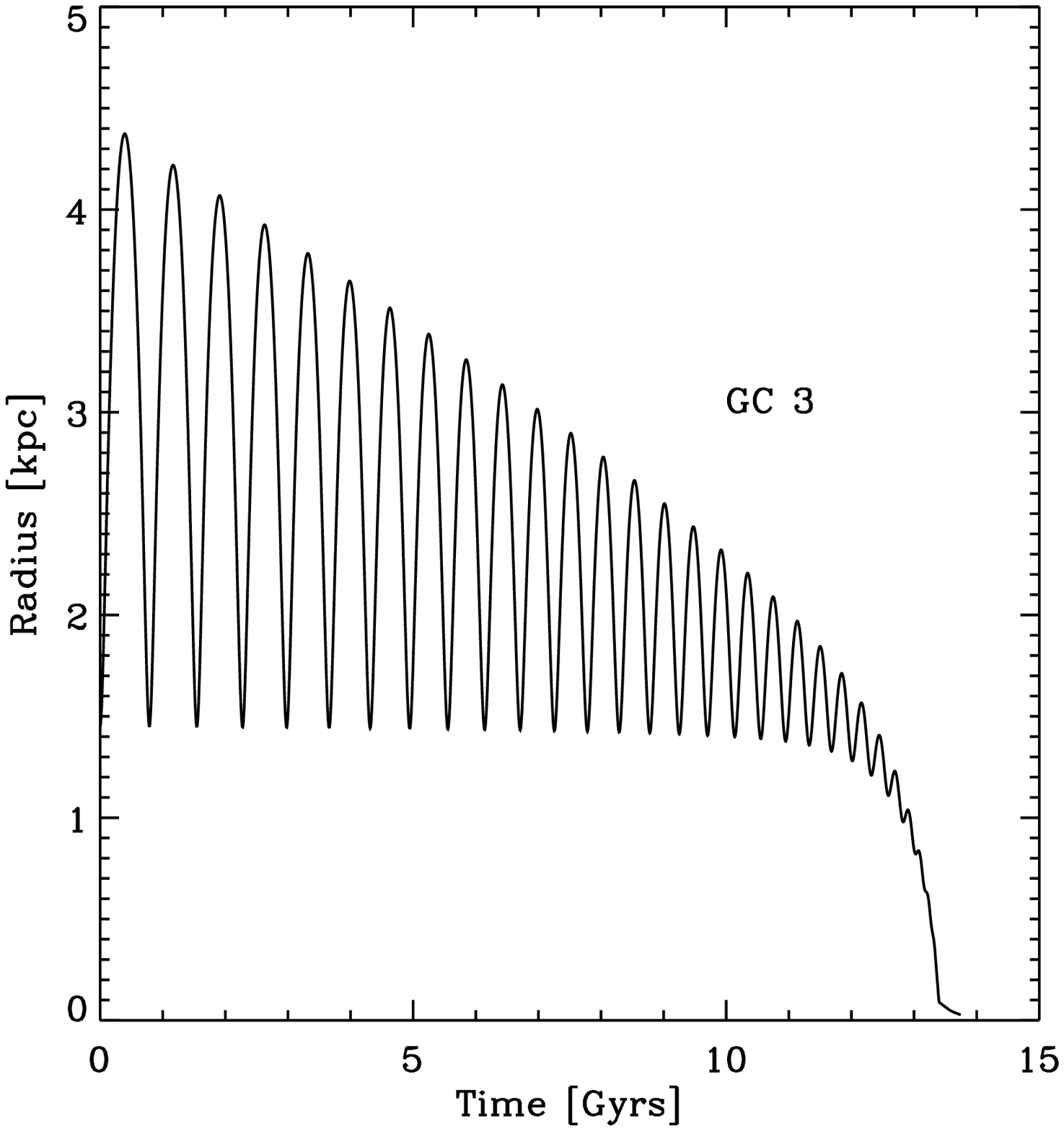}
}&
\subfigure{\label{fig:spi3}
\includegraphics[angle=0,width=7.0cm]{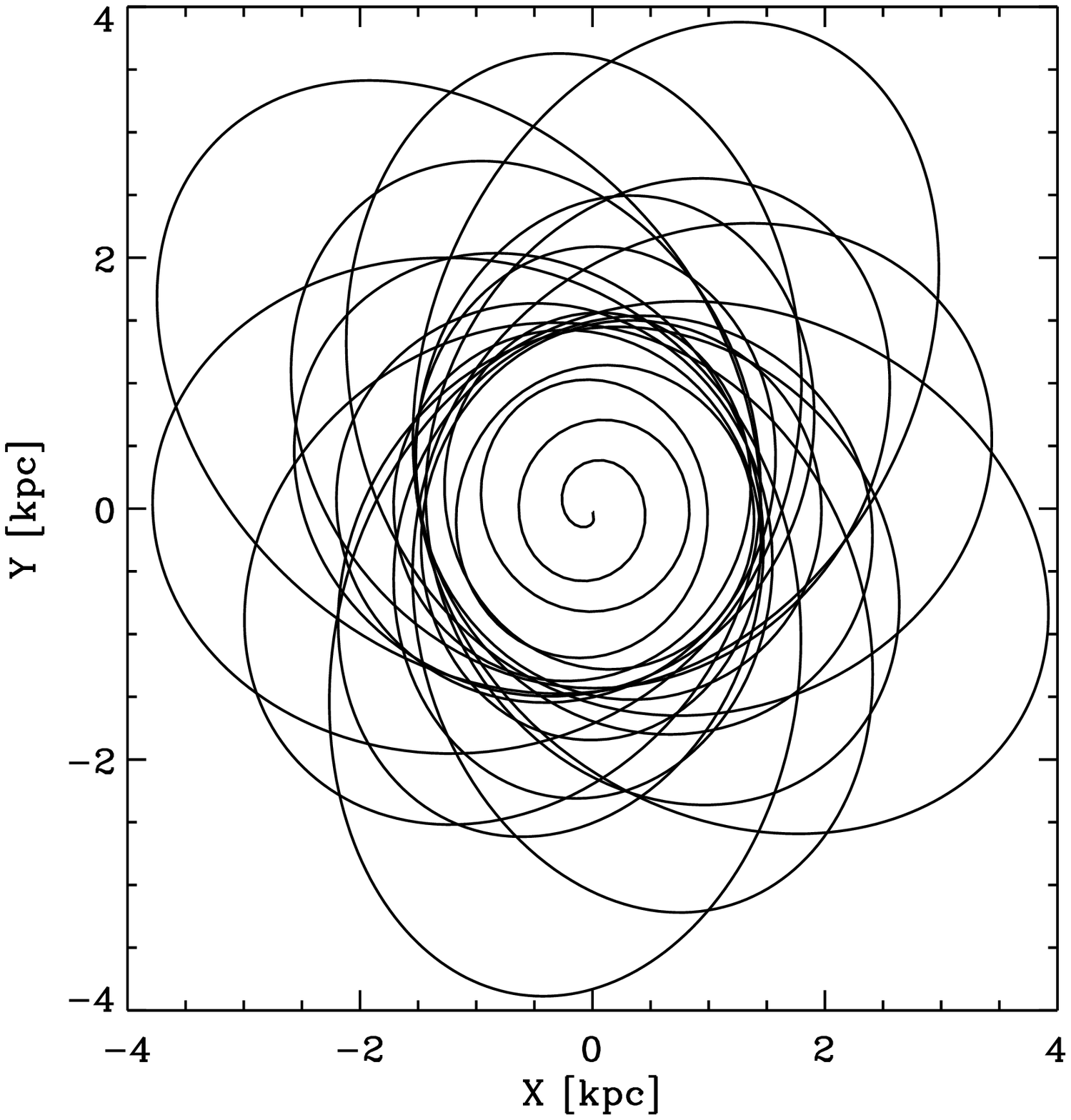}
}
\\
\subfigure{\label{fig:rad4}
\includegraphics[angle=0,width=7.0cm]{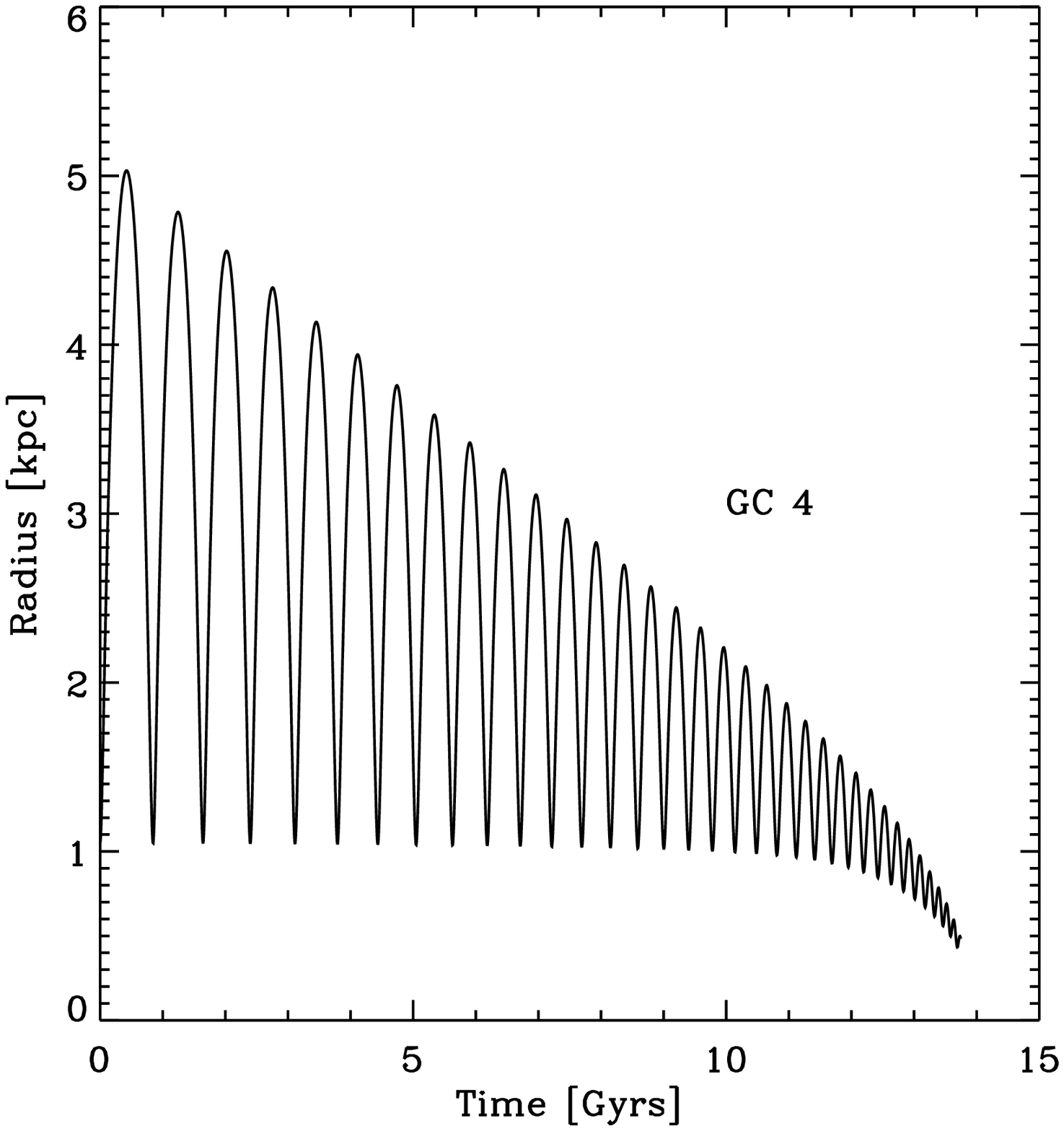}
}&
\subfigure{\label{fig:spi4}
\includegraphics[angle=0,width=7.0cm]{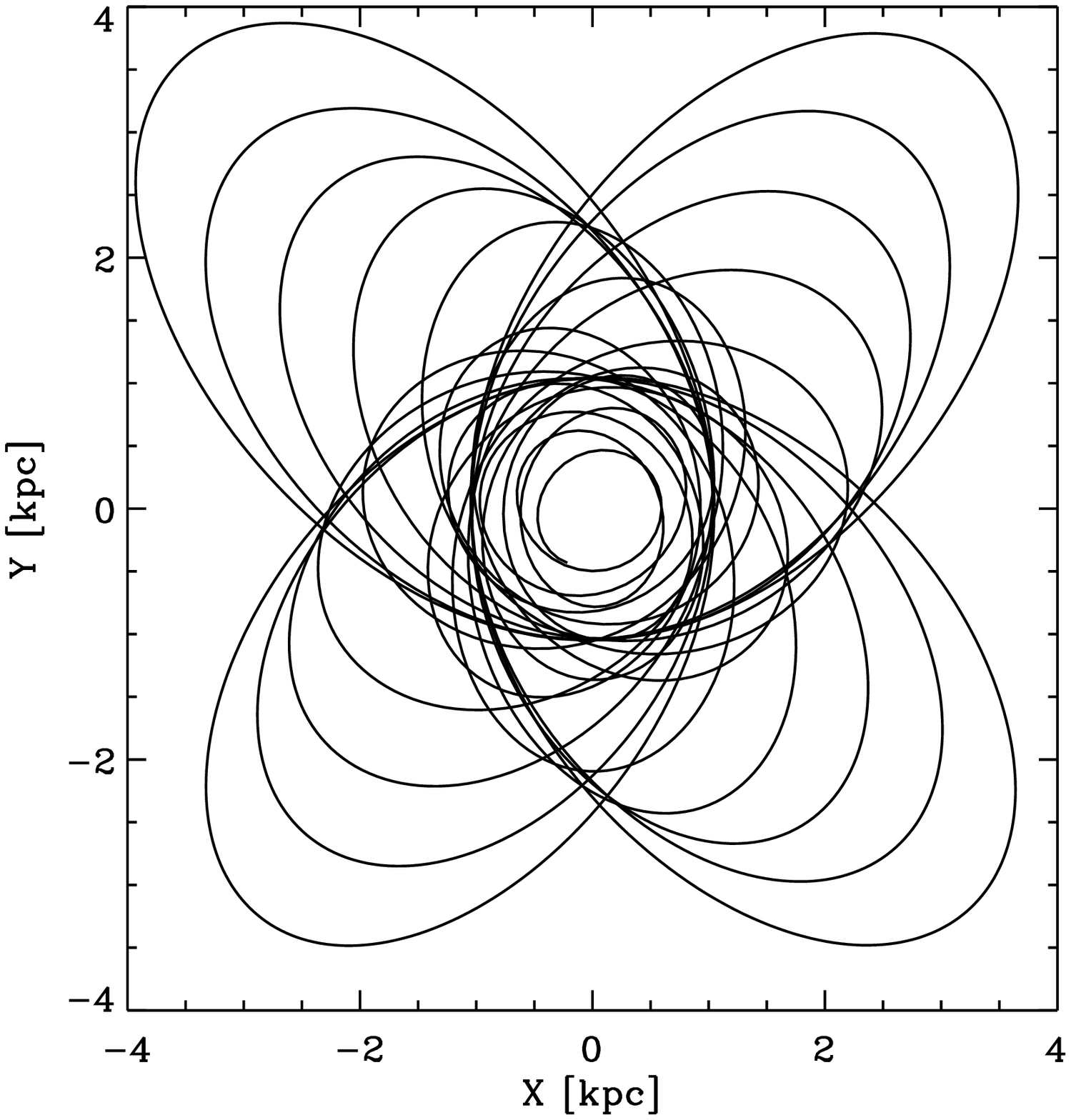}
}
\\
\end{tabular}

\caption{The orbital decay of highly radial orbits that enter well inside the tidal radius in MOND. The orbits decay in a Hubble time and spend a large fraction of the final Gyrs orbiting within the tidal radius. This scenario is discussed in depth in \S\ref{sec:radial}}
\label{fig:gcs}
\end{figure*}

One final point worth mentioning is that the timing problem is aggravated greatly by the existence of GC 3 (the most massive one by a factor of two). It is entirely possible that GC 3 orbits at a ``safe'' distance from Fornax, whereas the other four are not on orbits beyond the tidal radius but instead are gradually spiralling in. If this is the case then from Fig~\ref{fig:dfm} we know that the next two largest GCs (2 and 5) will take at least $8~Gyrs$ to decay. Comparing that to the first panel of Fig~\ref{fig:gcs}, to survive a Hubble time and plunge to within $1.2~kpc$ of Fornax the radial orbit must be relatively extreme. Recall that Fig~\ref{fig:gcs} was made to show the orbit which could plunge deepest into Fornax and still survive a Hubble time. If instead the orbit only skims the edge of Fornax, the radial orbit need not be as extreme. Thus, different orbits exist, both circular and exotic, which can survive in orbit of Fornax at the correct distance for a Hubble time.


\begin{figure}
\includegraphics[angle=0,width=7cm]{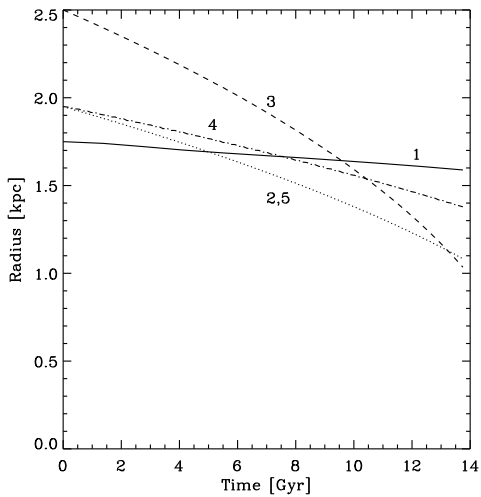}
\caption{The orbital decay of the 5 GCs for the DM profile beginning with circular orbits at radii that enable sensible current orbital radii.}\label{fig:dfn}
\end{figure}

\section{Discussion}
The fact that 5 GCs currently orbit the Fornax dwarf, although it has no stellar nucleus, would seem to suggest that dynamical friction is a weak means of draining their orbital energy. Significantly, the DM profile necessary to match the losVD profile of the Fornax dwarf can easily accommodate the slow decay of the GCs to orbital distances resembling their current ones (Fig~\ref{fig:dfn}).

Dynamical friction inside the tidal radius is far stronger in MOND (see Fig~\ref{fig:df}). GCs beginning on circular orbits narrowly inside the tidal radius lose their orbital energy to DF and sink to the centre of Fornax within a Hubble time. In the case of the largest GC, the orbit cannot be sustained for more than $5~Gyrs$. On the other hand, highly radial orbits that enter well inside the luminous part of the dwarf can survive for a Hubble time, with the final 2 or $3~Gyrs$ spent almost entirely within the luminous part before eventually spiralling in. Furthermore, if these GCs were captured in the last $4~Gyrs$, there is also not enough time to drain the angular momentum of even the most massive one. It must be cautioned, however, that these orbits are highly superficial and it seems highly unlikely that such extreme orbits exist at the expense of more pedestrian ones in every case, although it may be true for one.

Outside the tidal radius (at $1.9~kpc$) circular orbits can survive indefinitely, since there exist no stars to scatter and drag the GC. The only issue is the probability that the 4 large GCs can simultaneously have no projected radii greater than $1.4~kpc$, which we estimate to be greater than a third, which for a unique system is credible. A final approach is to ignore the largest GC which greatly relieves the timing problem, since the other four GCs can withstand the DF for substantially longer interior to the tidal radius.

Taking into account the discussion above and the uncertainties of the analytical approach to DF in MOND, it is probably not worth attaching too much significance to the existence of these 5 GCs. The biggest mystery for the DM paradigm is why this was seen as a critical problem in the first place.

\section{acknowledgments} GWA's research is supported by the University of Torino and Regione Piemonte. Partial support from the INFN grant PD51 and hospitality at the University of St. Andrews are also greatly appreciated. Support from the PRIN2006 grant ``Co\-stituenti fondamentali dell'Universo'' of the Italian Ministry of University and Scientific Research is gratefully acknowledged by AD.

\end{document}